# Sequential Nonparametric Regression

**Haijie Gu**  HAIJIEG@CS.CMU.EDU

Carnegie Mellon University, 5000 Forbes Ave., Pittsburgh, PA 15213 USA

**John Lafferty**  LAFFERTY@GALTON.UCHICAGO.EDU

University of Chicago, 5734 S. University Ave., Chicago, IL 60637 USA

## Abstract

We present algorithms for nonparametric regression in settings where the data are obtained sequentially. While traditional estimators select bandwidths that depend upon the sample size, for sequential data the effective sample size is dynamically changing. We propose a linear time algorithm that adjusts the bandwidth for each new data point, and show that the estimator achieves the optimal minimax rate of convergence. We also propose the use of online expert mixing algorithms to adapt to unknown smoothness of the regression function. We provide simulations that confirm the theoretical results, and demonstrate the effectiveness of the methods.

## 1. Introduction

Bandwidth selection is arguably the most important aspect of nonparametric regression using smoothing kernels. It is well understood how the optimal bandwidth for regression depends on the sample size. Considering the one dimensional case, let

$$Y_i = m(X_i) + \epsilon_i, \quad i = 1, \ldots, n \qquad (1.1)$$

where the $X_i$s are independent and identically distributed, $m : \mathbb{R} \to \mathbb{R}$ is the unknown function to estimate, and $\epsilon_i \sim N(0, \sigma^2)$. Assuming $m''$ is absolutely continuous and $\int m''(x)^2 dx < \infty$, the risk of the Nadaraya-Watson kernel regression estimator with bandwidth $h$ has the form

$$R(\widehat{m}_n, m) = c_1 h^4 + \frac{c_2}{nh} + o(nh^{-1}) + o(h^4) \qquad (1.2)$$

where

$$R(\widehat{m}_n, m) = \mathbb{E}_m \int (\widehat{m}_n(x) - m(x))^2 dx \qquad (1.3)$$

is defined as the risk of the estimate $\widehat{m}_n$ on a sample of size $n$. Here $c_1$ and $c_2$ are constants that depend on the kernel and the distribution of $X$. The optimal bandwidth that minimizes (1.2) thus has order $h^* = O(n^{-1/5})$, which leads to the optimal minimax rate of convergence $O(n^{-4/5})$; see Györfi et al. (2002). More generally, if we assume further smoothness of $m$ so that the $d$-th derivative of $m$ exists and is bounded, then the minimax rate $n^{-2d/(2d+1)}$ is achieved using local polynomial regression of order $d-1$ with a bandwidth of order $h^* = O(n^{-1/(2d+1)})$ (Fan & Gijbels, 1996). Thus, bandwidth selection is of the essence in nonparametric regression, and a large body of research has been devoted to this problem in various settings.

These classical results assume that a training data set of size $n$ is given; the sample size $n$ formally increases to infinity in the theoretical analysis. In an online setting, however, the data arrive sequentially, and the size of the data set is always changing. In this case the bandwidth needs to adapt somehow to the changing sample size. A simple variation of the classical methods would carry out a batch regression on the entire data set seen up to the current time $T$, with an appropriately sized bandwidth. However, this would require quadratic complexity $O(T^2)$ to compute the estimates after $T$ points are observed, and is prohibitive for large sample sizes. This motivates the problem studied in this paper—to develop computationally efficient estimators in the online setting that preserve the statistical efficiency of the classical batch estimators. We propose an algorithm for sequential regression that requires linear computational cost, and prove that the algorithm achieves the optimal minimax rate of convergence. The essential idea is to avoid recomputation by shrinking the bandwidth for each new observed data point. Our analysis assumes that the data are iid, and the algorithm assumes they are obtained sequentially.

Although online local polynomial smoothing has not been previously studied, significant previous work has been devoted to related problems. Steland (2010) investigates a cross-validation scheme for sequential data and establishes theoretical results. However, this does not consider the cost of recomputing the entire model for each new bandwidth; performing leave-one-out cross-validation adds extra computation and would be impractical for many online settings. Kivinen et al. (2004) develop variants of stochastic gradient descent for online learning in a reproducing kernel Hilbert space. The algorithms require linear computation cost at each step, and their RKHS analysis does not consider selection of tuning parameters for the kernel, or adaptation to the unknown smoothness of the regression function. A great deal of work has been carried out for the problem of adaptation in the classical batch setting. Fan & Gijbels (1995) consider using Residual Squares Criteria (RCS) for performing data-driven bandwidth selection; Ruppert et al. (1995) propose a plug-in bandwidth selection scheme for local linear kernel estimators. These are effective methods for adaptive estimation; however, they do not take into account the computational cost for online updating. The mixing expert framework has been a popular strategy for online prediction, and there is a rich literature on this topic (Cesa-Bianchi & Lugosi, 2006). Results by Bunea & Nobel (2008) give oracle inequalities for regression in terms of generalized simplex combinations of a set of fixed estimators in the online setting. However, this analysis does not allow the experts to change over time, as we require with dynamically changing bandwidths. Further related work is discussed below.

In the following section we present the algorithm for sequential local polynomial regression. In Section 3 we outline a theoretical analysis of the risk achieved by this algorithm for both sequential density estimation (Theorem 3.2) and regression (Theorem 3.3). In Section 4 we briefly discuss the problem of adapting to unknown smoothness, using the expert mixing framework and also the extension to additive models. In Section 5 we present experimental results showing that our algorithm is comparable to the batch algorithm but much more computationally efficient, and adapts to unknown smoothness of the true function.

## 2. Sequential Local Polynomial Smoothing

Our efficient sequential estimator is based on local polynomial regression with a sequence of shrinking bandwidths. Among the various nonparametric regression methods, local polynomial regression enjoys strong minimax properties, and gracefully deals with the problem of boundary bias (Fan et al., 1993). Let $\{(X_1, Y_1), (X_2, Y_2), \ldots\}$ be an observation sequence according to the model given in equation (1.1). We assume that the pairs $(X_i, Y_i)$ are independent and identically distributed random variables. Throughout we assume for simplicity that the domain of $m(x)$ is $x \in [0, 1]$. An extension to higher dimensional $x$ is straightforward. We discuss an extension to additive models for the high dimensional case in Section 4.2.

### 2.1. Sequential kernel regression

We first present the simplest version of the method. Recall that the kernel regression (Nadaraya-Watson) estimator in the batch setting is

$$\widehat{m}_n(x_0) = \frac{\sum_{i=1}^n K_h(X_i, x_0) Y_i}{\sum_{i=1}^n K_h(X_i, x_0)} \quad (2.1)$$

where $h$ is the bandwidth. Under standard assumptions that the regression function is in a second-order Sobolev space, the bandwidth is chosen as $h = c \cdot n^{-1/5}$, and the estimator achieves the minimax rate of convergence $R(\widehat{m}_n, m) = O(n^{-4/5})$. Now, suppose the data arrive sequentially. Our sequential estimator takes the form

$$\widetilde{m}_n(x_0) = \frac{\sum_{t=1}^n K_{h_t}(X_t, x_0) Y_t}{\sum_{t=1}^n K_{h_t}(X_t, x_0)} \quad (2.2)$$

where the bandwidth $h_t = c \cdot t^{-1/5}$ is used only for the $t$th point. The algorithm incrementally computes the numerator and denominator, shrinking the bandwidth for each new term added. A corollary of our main technical result is that this estimator achieves the same rate of convergence as the Nadaraya-Watson estimator.

This result is surprising since the first examples in the sequence, $(X_1, Y_1), (X_2, Y_2), \ldots$, are assigned bandwidths $h_t = c \cdot t^{-1/5}$ that are too large, introducing a bias in the estimate. However, our analysis shows that this bias is asymptotically "washed out." To gain some intuition for how this happens, note that we can write the estimate as

$$\widetilde{m}_n(x_0) = \frac{\frac{1}{n}\sum_{t=1}^n K_{h_t}(X_t, x_0) Y_t}{\frac{1}{n}\sum_{t=1}^n K_{h_t}(X_t, x_0)}. \quad (2.3)$$

Define $f : \mathbb{R} \to [0, 1]$ to be the marginal probability density function of $X$. As we show below, the denominator is a consistent density estimate of $f(x_0)$, and moreover it attains the optimal rate of convergence. The early contributions to the series $K_{h_1}(X_1, x_0) Y_1 + K_{h_2}(X_1, x_0) Y_2 + \cdots$ are then effectively weighted by $1/n\widehat{f}(x_0)$, which removes the bias they introduce as $n$ increases.

## 2.2. Sequential local polynomial smoothing

In the classical batch estimation case, the order-$d$ local polynomial regression at a prediction point $x_0$ is computed by minimizing the locally weighted squared error

$$\sum_{t=1}^{n} K\left(\frac{X_t - x_0}{h_n}\right) \left(Y_t - \sum_{j=0}^{d} \beta_j(x_0)(X_t - x_0)^j\right)^2 \quad (2.4)$$

where $K(\cdot)$ is the kernel. Denote by $\widehat{\boldsymbol{\beta}}(x_0)$ the $(d+1)$-vector that minimizes this objective. The regression estimate is then given by the intercept $\widehat{m}(x_0) = \widehat{\beta}_0(x_0)$. Let $\boldsymbol{X}_n$ be the $n \times (d+1)$ design matrix

$$\boldsymbol{X}_n = \begin{pmatrix} 1 & (X_1 - x_0) & \cdots & (X_1 - x_0)^d \\ \vdots & \vdots & \vdots & \vdots \\ 1 & (X_n - x_0) & \cdots & (X_n - x_0)^d \end{pmatrix}.$$

Furthermore, let $\boldsymbol{y}_n = (Y_1, Y_2, \ldots, Y_n)^T$, and let $\boldsymbol{W}_n = \text{diag}\{K_{h_n}(X_t, x_0)\}_{1 \leq t \leq n}$ be an $n \times n$ diagonal matrix of weights, where $K_h(X_t, x) = \frac{1}{h}K(\frac{X_t - x}{h})$. Then the solution that minimizes (2.4) is given as

$$\widehat{\boldsymbol{\beta}}(x_0) = (\boldsymbol{X}_n^T \boldsymbol{W}_n \boldsymbol{X}_n)^{-1} \boldsymbol{X}_n^T \boldsymbol{W}_n \boldsymbol{y}_n, \quad (2.5)$$

and $\widehat{m}(x_0) = \widehat{\beta}_0(x_0)$ is an estimate of the regression function at $x_0$ (Fan et al., 1993).

If we were to adapt $h_{n+1}$ to the increased sample size $n+1$ at the next time step, we would have to recompute the entire $X^T W X$ matrix. To save computation, we allow a variable bandwidth in $W$. In other words, we select a new bandwidth $h_{n+1}$ that only applies to $(X_{n+1}, Y_{n+1})$.

Specifically, we propose the following sequential local polynomial regression algorithm. Let

$$\widetilde{\boldsymbol{W}}_n = \text{diag}\{K_{h_t}(X_t, x_0)\}_{1 \leq t \leq n}$$

where $h_t = c \cdot t^{-1/(2d+1)}$ is the bandwidth that would be asymptotically optimal with respect to a sample of size $t$, with $c$ a constant. Our proposed online estimate after $n$ samples are observed is then

$$\widetilde{m}_n(x_0) = \boldsymbol{e}_1^T (\boldsymbol{X}_n^T \widetilde{\boldsymbol{W}}_n \boldsymbol{X}_n)^{-1} \boldsymbol{X}_n^T \widetilde{\boldsymbol{W}}_n \boldsymbol{y}_n. \quad (2.6)$$

where $\boldsymbol{e}_1 = (1, 0, \ldots, 0)$. Denote by $\boldsymbol{S}_n$ the $(d+1) \times (d+1)$ matrix $\boldsymbol{X}^T \widetilde{\boldsymbol{W}}_n \boldsymbol{X}$, with $(i, j)$ entry

$$S_n(i, j) = \sum_{t=1}^{n} K_{h_t}(X_t, x_0)(X_t - x_0)^{i+j}. \quad (2.7)$$

To update the model after having observed $(X_{n+1}, Y_{n+1})$, note that

$$\boldsymbol{S}_{n+1} = \boldsymbol{S}_n + K_{h_{n+1}}(X_{n+1}, x_0) \boldsymbol{x}_{n+1} \boldsymbol{x}_{n+1}^T \quad (2.8)$$

where $\boldsymbol{x}_{n+1}$ is the $(d+1)$ vector $(1, (X_{n+1} - x_0), (X_{n+1} - x_0)^2, \ldots, (X_{n+1} - x_0)^d)$. Now, using (2.8) and the Woodbury formula for the inverse of a rank-one matrix update,

$$(\boldsymbol{A} + \boldsymbol{v}\boldsymbol{v}^T)^{-1} = \boldsymbol{A}^{-1} - \boldsymbol{A}^{-1}\boldsymbol{v}(1 + \boldsymbol{v}^T \boldsymbol{A}^{-1} \boldsymbol{v})^{-1}\boldsymbol{v}^T \boldsymbol{A}^{-1}, \quad (2.9)$$

we have that updating $\boldsymbol{S}_{n+1}^{-1}$ given $\boldsymbol{S}_n^{-1}$ has complexity $O(d^2)$. Similarly, updating

$$\begin{aligned}\boldsymbol{X}_{n+1}^T \widetilde{\boldsymbol{W}}_{n+1} \boldsymbol{y}_{n+1} \\ = \boldsymbol{X}_n^T \widetilde{\boldsymbol{W}}_n \boldsymbol{y}_n + K_{h_{n+1}}(X_{n+1}, x_0)\, \boldsymbol{x}_{n+1} Y_{n+1}\end{aligned} \quad (2.10)$$

has $O(d)$ cost. Therefore, the complexity of updating the estimate from a sample of size $n$ to one of size $n+1$ at each point $x_0$ in a grid $\mathcal{G}$ of size $|\mathcal{G}|$ is $O(|\mathcal{G}|d^2)$, independent of $n$. However, the update cost in the batch setting is $O(|\mathcal{G}|(nd^2 + d^3))$ because changing the bandwidth forces re-evaluating equation (2.5).

## 3. Risk Analysis

In this section, we give a risk analysis of sequential density estimation and regression. Assuming the regression function $m$ lies in $C^d$, the class of functions with $d$ continuous derivatives, our goal is to show that the asymptotic risk of the online algorithm given in the previous section achieves the statistical rate of convergence of $n^{-2d/(2d+1)}$. This is the minimax optimal rate for this function class.

We begin by analyzing the risk of sequential kernel density estimation and kernel regression ($d = 2$), because the analysis is simpler and more transparent for these special cases than for the general case. We then generalize the results to order $d-1$ sequential polynomial regression in $C^d$.

We assume that the true density function $f$ and the true regression function $m$ have $d \geq 2$ continuous derivatives. The kernel $K$ is assumed to satisfy the following properties: $\int K(u)\,du = 1$, $\int K(u)\,u\,du = 0$, $\sigma_K^2 \equiv \int K(u)\,u^2 du < \infty$. We restrict the sequence of bandwidths $\{h_t \mid t = 1, 2, 3, \ldots\}$ to satisfy $\lim_{t \to \infty} h_t = 0$ and $\lim_{n \to \infty} \frac{1}{n^2} \sum_{t=1}^{n} \frac{1}{h_t} = 0$.

### 3.1. Sequential Kernel Density Estimation

Our sequential kernel density estimator $\widetilde{f}$ is given by $\widetilde{f}_n(x) = \frac{1}{n} \sum_{t=1}^{n} \frac{1}{h_t} K\left(\frac{x - X_t}{h_t}\right)$. This is computed incrementally according to the update rule

$$\widetilde{f}_{n+1}(x) = \frac{n}{n+1}\widetilde{f}_n(x) + \frac{1}{(n+1)h_{n+1}} K\left(\frac{x - X_{n+1}}{h_{n+1}}\right).$$

Thus, updating $\widetilde{f}_n$ has cost $O(|\mathcal{G}|)$, where $|\mathcal{G}|$ is the size of the grid of $x$ values at which the estimates are made.

**Lemma 3.1.** *The risk of the sequential density estimate $\widetilde{f}_n$ at time $t = n$ is*

$$R(f, \widetilde{f}_n) = \frac{1}{4} \int f''(x)^2 \, dx \left( \sigma_K^2 \frac{1}{n} \sum_{t=1}^n h_t^2 \right)^2 \quad (3.1)$$

$$+ \left( \frac{\sum_{t=1}^n \frac{1}{h_t}}{n^2} \right) \int K^2(u) \, du$$

$$+ o\left( \frac{(\sum_{t=1}^n h_t^2)^2}{n^2} \right) + o\left( \frac{\sum_{t=1}^n \frac{1}{h_t}}{n^2} \right). \quad (3.2)$$

The proof of Lemma 3.1 is given in Section A.1. We note in passing that a similar algorithm for sequential density estimation appears, without theoretical analysis, in Kristan et al. (2010).

**Theorem 3.1.** *Let $h_t = c\,t^{-1/5}$ for some constant $c$. Then the risk of sequential kernel density estimator $\widetilde{f}_n$ satisfies $R(f, \widetilde{f}_n) = O(n^{-4/5})$.*

*Proof.* Let $c_1 = \frac{1}{4}(\sigma_K^2)^2 \int f''(x)^2 dx$, and $c_2 = \int K^2(u)\,du$. Then Lemma 3.1 states that

$$R(f, \widetilde{f}_n) = \frac{1}{n^2} \left[ c_1 \left( \sum_{t=1}^n h_t^2 \right)^2 + c_2 \sum_{t=1}^n \frac{1}{h_t} \right]. \quad (3.3)$$

Let $h_t = c\,t^{-k}$ for $0 < k < \frac{1}{2}$. Then with this choice of bandwidth sequence

$$R(f, \widetilde{f}_n) = \frac{1}{n^2} \left[ c_1 \left( \sum_{t=1}^n t^{-2k} \right)^2 + c_2 \sum_{t=1}^n t^k \right] \quad (3.4)$$

$$\leq \frac{1}{n^2} \left[ c_1 \left( \int_0^n t^{-2k} dt \right)^2 + c_2 n^{k+1} \right] \quad (3.5)$$

$$= c_1 \frac{1}{(1-2k)^2} n^{-4k} + c_2 n^{k-1}. \quad (3.6)$$

Minimizing over $k$, we find that $k^* = \frac{1}{5}$, and optimal risk is of order $R^* = O(n^{-4/5})$. □

The above analysis assumes that the density $f$ has a continuous second derivative. The result can be extended to the case where $f$ is in $C^d$, using a higher order kernel satisfying $\sigma_K^j = \int u^j K(u)\,du = 0$ for all $j < d$.

**Theorem 3.2.** *If the density function $f$ lies in $C^d$, then the optimal risk of the sequential kernel density estimator $\widetilde{f}_n$ satisfies $R^* = O(n^{-2d/2d+1})$ when the bandwidth sequence is taken to be $h_t^* = c\,t^{-1/2d+1}$.*

*Proof sketch.* By a Taylor expansion and calculations similar to those in the proof of Lemma 3.1, we have

$$\mathbb{E}\widetilde{f}_n(x) = \frac{1}{n} \sum_{t=1}^n \int K(u) f(x - h_t u) du \quad (3.7)$$

$$= \frac{1}{n} \sum_{t=1}^n \left[ f(x) + \frac{h_t^2}{2} f''(x) \sigma_K^2 + \cdots + \frac{h_t^d}{d!} f^d(x) \sigma_K^d \right]$$

$$+ o(h_t^d).$$

Assuming a higher order kernel, we have $\sigma_K^j = 0$ for $j < d$, so the leading order of the bias is

$$\mathrm{Bias}(\widetilde{f}_n(x)) = \frac{1}{d!} f^d(x) \sigma_K^d \left( \frac{\sum_{t=1}^n h_i^d}{n} \right) + o\left( \frac{\sum_{t=1}^n h_i^d}{n} \right).$$

The variance satisfies $\mathrm{Var}(\widetilde{f}_n(x)) = f(x) \int K^2(u) du \cdot \left( \frac{1}{n^2} \sum_{t=1}^n \frac{1}{h_t} \right) + o\left( \frac{1}{n^2} \sum_{t=1}^n \frac{1}{h_t} \right)$ as shown in the proof of Lemma 3.1. Thus, using the bias-variance decomposition the risk takes the form

$$c_1 \left( \frac{1}{n} \sum_{t=1}^n h_t^d \right)^2 + c_2 \frac{1}{n^2} \sum_{t=1}^n \frac{1}{h_t}. \quad (3.8)$$

Assuming that $h_t = c\,t^{-k}$ for $k > 0$ and optimizing over $k$ yields $k^* = \frac{1}{2d+1}$, bandwidth sequence $h_t^* = c\,t^{-1/2d+1}$, and optimal risk $R^* = O(n^{-2d/2d+1})$. □

### 3.2. Sequential Local Polynomial Regression

Similar results hold for the sequential local polynomial regression estimator. In particular, for $d = 2$ we use

$$\widetilde{m}_n(x) = \frac{1}{n\widetilde{f}_n(x)} \sum_{t=1}^n K_{h_t}(x, X_t) Y_t \quad (3.9)$$

where $\widetilde{f}_n(x) = \frac{1}{n} \sum_{t=1}^n K_{h_t}(x, X_t)$. The following result is proved in Section A.1.

**Lemma 3.2.** *The risk of the estimator $\widetilde{m}_n(x)$ in (3.9) satisfies*

$$R(\widetilde{m}_n, m) \quad (3.10)$$

$$= \frac{1}{4} \left( \sigma_K^2 \frac{1}{n} \sum_{t=1}^n h_t^2 \right)^2 \int \left( m''(x) + 2m'(x) \frac{f'(x)}{f(x)} \right)^2 dx$$

$$+ \sigma^2 \int K^2(x)\,dx \int \frac{dx}{f(x)} \left( \frac{1}{n^2} \sum_{t=1}^n \frac{1}{h_t} \right)$$

$$+ o\left( \frac{1}{n} \sum_{t=1}^n h_t^2 \right)^2 + o\left( \frac{1}{n^2} \sum_{t=1}^n \frac{1}{h_t} \right).$$

The analogue of Theorem 3.2 for density estimation can also be obtained in the regression setting. Instead of choosing a special kernel to cancel out the lower order terms in the Taylor's series, we leverage the minimax optimality of local polynomial regression as introduced in Section 2.

**Theorem 3.3.** *Suppose that the regression function $m$ has $d$ continuous derivatives. Let $h_t = c\, t^{-1/2d+1}$. Then at $t = n$, the order $d-1$ sequential local polynomial regression attains the optimal minimax risk $R^* = O(n^{-2d/2d+1})$.*

The proof of Theorem 3.3 is given in Section A.3.

## 4. Extensions

In this section we extend the above analysis in two ways. First, we show how the expert mixing framework can be used to adapt to the smoothness exponent. Second, we show how the procedure can be extended to additive models.

### 4.1. Adapting to unknown smoothness

The theoretical performance of the sequential estimators presented above hinges on selecting the correct order $d$ of the local polynomial, and in practice it depends on the constant $c$ in the bandwidth as well. Traditional statistical model selection methods, e.g. AIC and cross validation, are not practical in an online scenario.

In order to maintain a reasonable computational cost, we combine estimators that use different parameters (order $d$ and constant $c$) through an exponential weighting strategy. Leveraging our analysis in Section 3, it can be shown that the procedure adapts to the unknown smoothness at the optimal rate, while maintaining a linear computational cost.

In more detail, our mixing online regression estimates procedure forms an exponential weighting of a set of sequential local polynomial regression estimates with different orders $d$ and bandwidth constants $c$.

Let $\mathcal{C} = \{c_i\}_{1 \leq i \leq C}$, and $\mathcal{D} = \{d_j\}_{1 \leq j \leq D}$. Define $M = CD$ to be the size of the family of sequential regression estimates $\mathcal{M} = \{\widetilde{m}_{(i,j),t}\}$ where $1 \leq i \leq C$ and $1 \leq j \leq D$. At time $t$, the bandwidth of the regression $\widetilde{m}_{(i,j),t}$ is given by $h_{(i,j),t} = c_i \cdot t^{-1/2d_j+1}$, and the corresponding estimator is computed as in equation (2.6); specifically,

$$\widetilde{m}_{(i,j),t}(x_0) = \boldsymbol{e}_1^T (\boldsymbol{X}_t^T \widetilde{\boldsymbol{W}}_{(i,j),t} \boldsymbol{X}_t)^{-1} \boldsymbol{X}_t^T \widetilde{\boldsymbol{W}}_{(i,j),t} \boldsymbol{y}_t \tag{4.1}$$

where

$$\widetilde{\boldsymbol{W}}_{(i,j),t} = \operatorname{diag}\{K_{c_i s^{-1/2d_j+1}}(X_s, x_0)\}_{1 \leq s \leq t}.$$

The double index $(i,j)$ is used to illustrate the construction of the expert set; in the following, for simplicity, we use a single index $k$ to index the $M$ sequential estimators.

Let $\ell_{k,t} = \sum_{s=1}^t (Y_s - \widetilde{m}_{k,s-1}(X_s))^2$ be the cumulative loss of estimator $k$ at time $t$. At each time $s$, the prediction of $Y_s$ is made with the estimator $\widetilde{m}_{k,s-1}$ constructed using the previous data $\{(X_1, Y_1), (X_2, Y_2), \ldots, (X_{s-1}, Y_{s-1})\}$. Let $w_{k,0} = M^{-1}$, and for $t \geq 1$ let

$$w_{k,t} = \frac{\exp(-\eta \ell_{k,t})}{\sum_{k'=1}^M \exp(-\eta \ell_{k',t})} \tag{4.2}$$

where $\eta$ is a positive learning rate, to be chosen later. The combined estimator $\widetilde{m}_*$ at time $t$ is the convex combination given by

$$\widetilde{m}_{*,t}(x_0) = \sum_{k=1}^M w_{k,t} \widetilde{m}_{k,t}(x_0). \tag{4.3}$$

Note that the weight at time $t$ can be updated in linear time $O(M)$ using

$$w_{k,t} = \frac{w_{k,t-1} \exp\left\{-\eta(Y_t - \widetilde{m}_{k,t-1}(X_t))^2\right\}}{\sum_{k'=1}^K w_{k',t-1} \exp\left\{-\eta(Y_t - \widetilde{m}_{k',t-1}(X_t))^2\right\}}. \tag{4.4}$$

A large literature is devoted to the study of regret-based bounds for online learning, which hold for any realization of data (Cesa-Bianchi & Lugosi, 2006). However, we are primarily interested in analyzing the statistical risk of our estimators. In particular, it is of interest to obtain an oracle inequality of the form

$$\mathbb{E}\|\widetilde{m}_{*,n} - m\|^2 \leq \min_{k=1,\ldots,M} \mathbb{E}\|\widetilde{m}_{k,n} - m\|^2 + \delta(n) \tag{4.5}$$

where $\delta(n)$ is the additive penalty paid for adaptation. Having already established the minimax rate of the oracle $\widetilde{m}_{\star,n} = \arg\min_{k=1,\ldots,M} \mathbb{E}\|\widetilde{m}_{k,n} - m\|^2$, this will enable us to show that the weighted combination adapts to achieve a (near) minimax rate.

**Theorem 4.1.** *Let $m \in C^{d^\star}$. Assume that*

1. $\widetilde{m}_{k,n} \in [-A, A]$, *for some constant $A$, for all $k, n$;*

2. $\min_k \mathbb{E} \exp |\widetilde{m}_{k,n} - Y_n| \leq L$, *for some $0 < L < \infty$.*

Suppose that the optimal order $d^\star$ is contained in the set of candidate degrees $\mathcal{D}$. Then, for sufficiently large $n$,

$$\mathbb{E}\|\widetilde{m}_{*,n} - m\|^2 \leq \min_{k=1,\ldots,K} \mathbb{E}\|\widetilde{m}_{k,n} - m\|^2 + \frac{\ln M}{n\eta} \quad (4.6)$$

where $\eta$ is chosen to be a small constant depending on $A$. In particular, we have that $R(m, \widetilde{m}_{*,n}) = O\left(n^{-2d^\star/2d^\star+1}\right)$.

This result is proved by adapting analysis by Yang (2004) (Theorem 5, see also Catoni (2004)).

### 4.2. Additive models

It is also possible to derive a sequential extension to the backfitting algorithm for additive models (Hastie & Tibshirani, 1990). Assuming now that $X$ is $p$ dimensional, an additive regression model takes the form

$$Y_i = m_0 + \sum_{j=1}^{p} m_j(X_i^j) + \epsilon_i, \quad i = 1, \ldots, n \quad j = 1, \ldots, p$$

where $m_0$ is an overall mean or intercept, the nonparametric regression functions $m_j$ are one-dimensional, and $\epsilon_i$ is mean zero noise. The backfitting algorithm is a type of coordinate descent or Gauss-Seidel procedure that iteratively computes the residuals for a variable $j$, and then smoothes those residuals to get a nonparametric estimate of the component function.

In our sequential setting, a complication comes from the fact that the residuals must be updated sequentially as well—for computational efficiency we cannot afford to compute the residuals over all of the previous data. Since the residuals for all of the $p$ variables depend on one another, we update them iteratively using our sequential regression estimator until convergence. We then cycle through the variables to sequentially update each component function in terms of the converged residuals.

This algorithm is made explicit below, where $update(m_j, X, resid, t)$ updates the $j$th model $m_j$ with the $(X, resid)$ pair, using the appropriate bandwidth at time $t$. This can be viewed as an incremental version of the smoothing step in the classical backfitting algorithm and can be efficiently computed using (2.8) and (2.10) in the case of local-polynomial smoothing, using (2.3) in the case of kernel regression. We note that the numerical convergence of even the classical backfitting is difficult to analyze. Example simulations of our sequential backfitting algorithm are given in Section **??**, where it compares favorably to the classical batch backfitting algorithm.

**Input**: $(X_1, Y_1), \ldots, (X_n, Y_n), \ldots$ sequentially
**Output**: $\widehat{m}^1, \cdots, \widehat{m}^n \in \mathbb{R}^p$
Initialize mean $m_0 = 0$;
Initialize $m_j = array(0, |X^j|), j = 1, \cdots, p$;
**foreach** *timestep $t$* **do**
    $m_0 = \frac{t-1}{t} m_0 + \frac{Y_t}{t}$;
    Initialize $resid = array(0, p)$;
    **while** *resid not converged:* **do**
        **foreach** $j = 1, \cdots, p$ **do**
            **foreach** $k \neq j$ **do**
                $m_k' = update(m_k, X_t^k, resid[k], t)$;
                $m_k' = m_k' - mean(m_k')$ ;
            **end**
            $resid[j] = Y_t - m_0 - \sum_{k \neq j} m_k'(X_t^k)$;
        **end**
    **end**
    **for** $j$ *in* $1, \cdots, p$ **do**
        $m_j = update(m_j, X_t^j, resid[j], t)$;
        $m_j = m_j - mean(m_j)$ ;
    **end**
    Output: $\widehat{m}^t = [m_1 \cdots m_p]$;
**end**
**Algorithm 1:** Sequential Backfitting

## 5. Experiments

To illustrate the performance of a single online estimator (descirbed in §2), we compare it with the batch algorithm using the same bandwidth $h = cn^{-1/5}$. We consider the following 4 regression functions of different smoothness on $[0, 1]$ used by Yang (2001): $f_1(x) = \exp(-x)$, $f_2(x) = 1 + 2x^2 + \exp(-5(x-0.5)^2)$, $f_3(x) = 1 + 2x^2 + \exp(-200(x-0.5)^)$, $f_4(x) = \exp(-200(x-0.2)^2)/\sqrt{0.005\pi} + \exp(-200(x-0.8)^2)/\sqrt{0.005\pi}$. The sample size is 150 and $\sigma^2 = 0.5$.

We choose the bandwidth constant $c$ from $\mathcal{C} = \{0.05, 0.07, 0.1, 0.3, 0.5, 0.7, 1, 1.5\}$. In addition, we add a comparison to a batch estimator that does leave one out cross validation at each time step to choose the bandwidth constant from $\mathcal{C}$. The first 50 samples are used to initialize the batch estimators.

Figure 1 (first row) shows the average loss as a function of the sample size for the best sequential estimator (with optimal constant), the best batch estimator, and the batch cross-validation estimator for the functions. In each case, the best sequential estimator has very similar loss to the best batch estimator, which supports our theory showing that the sequential estimator achieves the same statistical rate of convergence as the classical batch algorithm. Note that in each case the average loss converges to around the noise level $\sigma^2 = .5$. The middle row compares the fits of the se-quential, batch, and batch with cross validation, and indicates that the fits are comparable, and very close to the true function. The third row shows the standard errors for the sequential estimators running the simulations 100 times for each function.

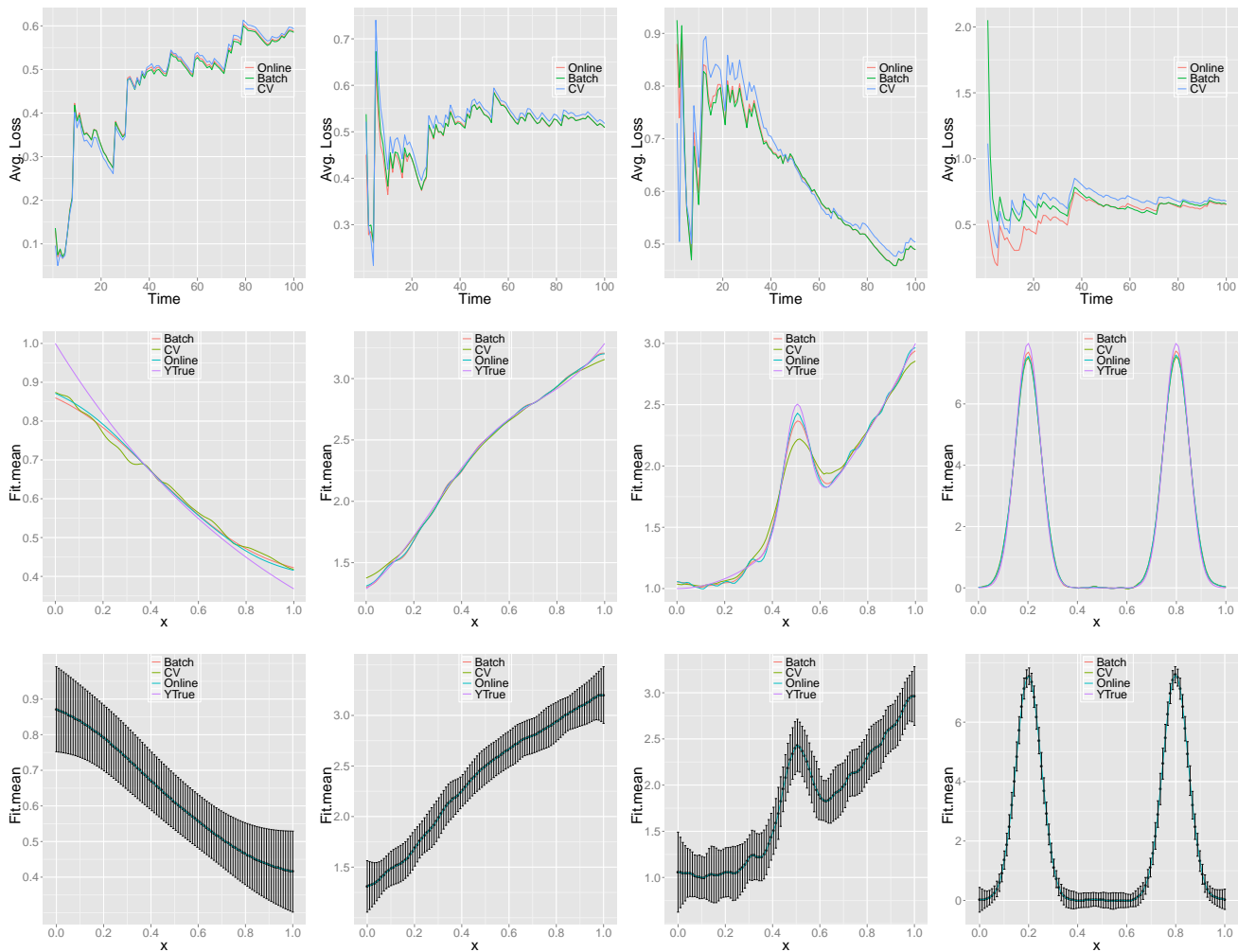

*Figure 1.* First row: average loss as a function of sample size $n$, for each of the four functions. The best sequential estimator is shown in red, the best batch estimator in green, and the cross-validation estimator in blue. Second row: fits for each method at $n = 150$. Bottom row, the mean and standard errors for the sequential estimators at $n = 150$.

Here we study the estimator's risk under unknown smoothness. We use the following parameterized set of functions: $m_\alpha(x) = 2|0.5(x - 0.5)|^\alpha, x \in [0,1]$ and let $\alpha = \{1, 1.5, 2, 2.5\}$. At the point $x_0 = 0.5$, the smoothness of the function $m_\alpha$ is Hölder $\alpha$ continuous. We consider four "experts" of online kernel estimators using the parameter set $\alpha$ for their bandwidths: $\widetilde{m}_\alpha$ takes the variable bandwidth $h_t = 0.4 \times t^{-1/(2\alpha+1)}$. We use the Gaussian kernel $K_h(u) = (2\pi)^{-1/2} e^{-u^2/(2h^2)}$ for the estimator, and set the noise to be $\sigma^2 = 0.01$. Although we have not included the analysis due to lack of space, it can be shown that our sequential estimator with shrinking bandwidths $c \cdot t^{-1/(2\alpha+1)}$ achieves the minimax rate $n^{-2/(2\alpha+1)}$ for this Hölder family of regression functions. The task here is to adapt to the unknown exponent $\alpha$.

Figure 2 shows the risk at the critical point $x_0 = 0.5$ of the four experts under the four functions, having different degrees of smoothness. We run simulations for each of the functions 1000 times, and show a plot of the average risk as a function of sample size. For each true function $m_\alpha$, the expert with the correct $\alpha$ obtains the lowest risk for any sample size. This suppports the analysis in §4, showing the best expert has the highest weight when the experts are mixed together. As a result, the mixing expert framework of online estimators adapts to the unknown smoothness of the true function. While the sequential algorithm makes a tradeoff between performance and computational efficiency, its performance is quite comparable to that of the optimal batch estimator.

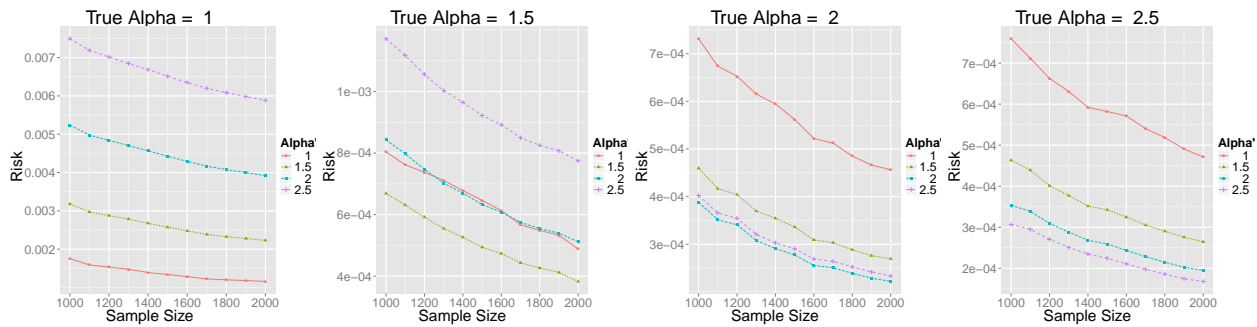

*Figure 2.* Risk (at critical point) vs. sample size for the four experts $\widetilde{m}_{\alpha'}$ under four functions $m_\alpha$ for (a) $m_1$, (b) $m_{1.5}$, (c) $m_2$ and (d) $m_{2.5}$. Experts $\widetilde{m}_1, \widetilde{m}_{1.5}, \widetilde{m}_2, \widetilde{m}_{2.5}$ are shown in red, green, blue and purple respectively. The risk is smallest for the expert with optimal order, showing that the algorithm adapts to the smoothness of the regression function.

## 6. Summary and Conclusions

We proposed and analyzed an efficient sequential algorithm for local polynomial smoothing in nonparametric regression. The first contribution of this work is the online algorithm that shrinks the bandwidth for each new point that arrives. The second is the analysis showing that order $d$ sequential local polynomial smoothing achieves the optimal minimax rate of convergence $n^{-2d/2d+1}$. Finally, we show that exponential weight mixing of a family of such sequential estimators adapts to unknown smoothness at the optimal rate, and extended the algorithm to sequential backfitting for nonparametric additive models. Our experimental results confirm the theoretical analysis, and show that little loss in statistical efficiency is sacrificed by the computationally efficient online procedure. While we have shown adaptation to global smoothness, an interesting direction for future work is to consider sequential estimators that adapt to spatially inhomogeneous function classes. One promising direction is to adapt the variable bandwidth estimator of Lepski et al. (1997) to online regression for Besov spaces.

## Acknowledgements

Research supported in part by NSF grant IIS-1116730 and AFOSR contract FA9550-09-1-0373.

## References

Bunea, Florentina and Nobel, Andrew. Sequential procedures for aggregating arbitrary estimators of a conditional mean. *Information Theory, IEEE Transactions on*, 54:1725–1735, 2008.

Catoni, Olivier. Statistical learning theory and stochastic optimization. In Picard, J. (ed.), *Ecole d'Eté des Probabilités de Saint Flour, XXXI–2001*, volume 1851. Springer, 2004.

Cesa-Bianchi, Nicolo and Lugosi, Gabor. *Prediction, Learning and Games*. Cambridge University Press, 2006.

Fan, Jianqing and Gijbels, Irene. Data-driven bandwidth selection in local polynomial fitting: variable bandwidth and spatial adaptation. *Journal of the Royal Statistical Society. Series B(Methodological)*, 57(2):371–394, 1995.

Fan, Jianqing and Gijbels, Irene. *Local Polynomial Modelling and Its Applications*, pp. 57–105. Chapman and Hall/CRC, 1996.

Fan, Jianqing, Gijbels, Irene, Gasser, Theo, Brockmann, Michael, and Engel, Joachim. Local polynomial fitting: A standard for nonparametric regression, 1993.

Györfi, Lásló, Kohler, Michael, Krzyżak, Adam, and Walk, Harro. *A Distribution-Free Theory of Nonparametric Regression*. Springer-Verlag, 2002.

Hastie, T. J. and Tibshirani, R. J. *Generalized additive models*. London: Chapman & Hall, 1990. ISBN 0412343908.

Kivinen, J., Smola, A. J., and Williamson, R. C. Online learning with kernels. *Signal Processing, IEEE Transactions on*, 52:2165–2176, 2004.

Kristan, M., Skocaj, D., and Leonardis, A. Online kernel density estimation for interactive learning. *Image and Vision Computing*, 28:1106–1116, 2010.

Lepski, O.V., Mammen, E., and Spokoiny, V.G. Optimal spatial adaptation to inhomogeneous smoothness: An approach based on kernel estimates with variable bandwidth selectors. *The Annals of Statistics*, 25(3):929–947, 1997.

Ruppert, D., Sheather, S. J, and Wand, M. P. An effective bandwidth selector for local least squares regression. *Journal of the American Statistical Association*, 90(432), 1995.

Steland, Ansgar. Sequential data-adaptive bandwidth selection by cross validation for nonparametric prediction, 2010.

Yang, Yuhong. Adaptive regression by mixing. *Journal of the American Statistical Association*, 96(454):574–588, 2001.

Yang, Yuhong. Combining forecasting procedures: Some theoretical results. *Econometric Theory*, 20:176–222, 2004.